\documentstyle[12pt]{article}
\newlength{\extraspace}
\newlength{\extraspaces}
\newcommand{\be}{\begin{equation}
\addtolength{\abovedisplayskip}{\extraspaces}
\addtolength{\belowdisplayskip}{\extraspaces}
\addtolength{\abovedisplayshortskip}{\extraspace}
\addtolength{\belowdisplayshortskip}{\extraspace}}
\newcommand{\ee}{\end{equation}}

\newcommand{\ba}{\begin{eqnarray}
\addtolength{\abovedisplayskip}{\extraspaces}
\addtolength{\belowdisplayskip}{\extraspaces}
\addtolength{\abovedisplayshortskip}{\extraspace}
\addtolength{\belowdisplayshortskip}{\extraspace}}
\newcommand{\ea}{\end{eqnarray}}

\newcommand{\A}{&\!\!\!}
\newcommand{\nonu}{\nonumber \\[.5mm]}

\newcommand{\VEV}[1]{\left\langle {#1} \right\rangle}

\begin{document}
\addtolength{\baselineskip}{.7mm}
\thispagestyle{empty}

\begin{flushright}
STUPP--00--159 \\ April, 2000
\end{flushright}
\vspace{.6cm}

\begin{center}
{\large{\bf{
Logarithmic Behaviours in the Feigin-Fuchs Construction \\[2mm]
of the $c=-2$ Conformal Field Theory
}}}
\\[20mm]

{\sc Hiroki Hata}
\footnote{\ E-mail: hata@krishna.th.phy.saitama-u.ac.jp} \\[7mm]

and \\[7mm]

{\sc Shun-ichi Yamaguchi}
\footnote{\ E-mail: yama@krishna.th.phy.saitama-u.ac.jp} \\[12mm]

{\it Department of Physics, Faculty of Science \\[2mm]
Saitama University, Urawa, Saitama 338-8570, Japan} \\[20mm]

{\bf Abstract} \\[7mm]

{\parbox{13cm}{\hspace{5mm}
We obtain logarithmic behaviours of a four-point correlation 
function in the $c=-2$ conformal field theory by using 
the Feigin-Fuchs construction. 
It becomes an indeterminate form by a naive evaluation, but is 
obtained by introducing an appropriate regularization procedure. 
}} 

\end{center}
\vfill
\newpage
\setcounter{section}{0}
\setcounter{equation}{0}

Conformal field theories whose correlation functions have 
logarithmic behaviour were first studied by Gurarie 
in the central charge $c=-2$ model \cite{LCFT1}. 
This model is one of the simplest systems 
since the correlation function in the problem consists of 
only one kind of primary fields. 
Logarithmic conformal field theories have interesting properties: 
new operators are needed, which are called logarithmic operators 
and never appeared for ordinary conformal field theories. 
The origin of the logarithms of Ref.\ \cite{LCFT1} is 
a hypergeometric function, which is a solution of 
the differential equation for the four-point correlation function 
and equivalent to the complete elliptic integral of the first kind. 
The origin should also be explained by the Feigin-Fuchs 
construction \cite{DF}, but the approach only to 
the logarithmic operators is studied \cite{KL}. 
The construction also applies to other models \cite{KAR}. 

In this paper 
the four-point correlation function of Ref.\ \cite{LCFT1} 
is calculated by using the Feigin-Fuchs construction. 
In this construction the correlation function 
is given by an integral representation. 
We find that its integral value becomes an indeterminate form, $\frac00$. 
In order to evaluate this form 
we introduce an appropriate regularization procedure: 
we perform an analytic continuation of a parameter in the hypergeometric 
function, namely, take the limit $c \rightarrow -2$ \cite{OTHERS}, 
and evaluate the form of the correlation function by 
using essentially the de l'Hospital theorem. 
In this way the logarithmic term appears, 
and our result is in agreement with that of Ref.\ \cite{LCFT1}. 
Application of our method to generic four-point correlation functions 
with logarithms is under studying. 

We now consider the Feigin-Fuchs construction \cite{DF} 
of conformal field theories. 
The action is given by 
\be
S = {1 \over 8\pi} \int d^2 \xi \sqrt{g}
\left(g^{\mu\nu} \partial_\mu \phi \partial_\nu \phi 
+2i\alpha_0 R \phi \right),
\label{action}
\ee
where $\phi$ is a real scalar field and $R$ is the scalar curvature 
on a sphere with fixed reference metric $g_{\mu\nu}$. 
The parameter $2\alpha_0$ can be interpreted as the background charge. 
On the complex plane, the energy momentum tensor is of the form 
\be
T(z) = -{1 \over 2} \partial_z \phi \partial_z \phi 
+i\alpha_0 \partial_z^{\,2} \phi, 
\label{EMT}
\ee
and two-point function of the field $\phi$ is 
$\VEV{\phi(z,\overline z)\phi(w,\overline w)} =-\ln|z-w|^2$. 
Thus the central charge of the system is written 
in terms of $\alpha_0$ as 
\be
c = 1-12 \alpha_0^{\,2}. 
\label{c}
\ee
When considering correlation functions of primary fields $\Phi$, 
we treat the correlation function in terms of the vertex operators 
$V_\alpha \equiv e^{i \alpha \phi}$ instead of $\Phi$. 
The conformal weight $h$ of the operator $V_{\alpha}$ is 
\be
h(V_{\alpha}) = h(V_{2\alpha_0 - \alpha}) = 
- \frac12 \alpha (2\alpha_0 - \alpha), 
\label{vpq}
\ee
so two admissible operators exist for one field $\Phi$. 
For the $(p,q)$ primary fields $\Phi_{p,q}$ \cite{BPZ}, 
the conformal weight takes the discrete value 
\be
h_{p,q} = -\frac12 \alpha_0^{\,2} 
+ \frac18 (p\alpha_+ + q\alpha_-)^2, 
\ee
where $\alpha_{\pm} = \alpha_0 \pm \sqrt{\alpha_0^{\,2} + 2}$. 
The corresponding operator $V_{\alpha}$ has 
the parameter $\alpha$ given by 
\be
\alpha_{p,q} = \alpha_0 
- \frac12 \left( p\alpha_+ + q\alpha_- \right). 
\ee
We also need the screening charges 
\be
Q_{\pm} = \int d^2u \, e^{i \alpha_{\pm} \phi(u,\overline u)}. 
\label{sc}
\ee
A certain number of screening charges should be inserted 
in the correlation function 
so that the charge neutrality condition, required by 
the zero mode integration of $\phi$, is satisfied.
Since the screening charges are the integrals of 
the operators with conformal weight one, 
the insertion have no effects on the conformal properties of 
correlation functions. 

Let us now concretely consider the four-point correlation function 
of Ref.\ \cite{LCFT1} 
\be
G^{(4)} \equiv 
\VEV{\mu(z_1,\overline z_1) \mu(z_2,\overline z_2)
\mu(z_3,\overline z_3) \mu(z_4,\overline z_4)} 
\label{G4}
\ee
in the $c=-2$ model. Here 
$\mu(z,\overline z) \equiv \Phi_{1,2}(z,\overline z)$ is 
the primary field with 
the conformal weight $h_{1,2}=-\frac18$. 
In the Feigin-Fuchs construction, 
four-point correlation functions of primary fields $\Phi_i$'s 
in general take the form \cite{DF} 
\be
\VEV{\Phi_1(z_1,\overline z_1) \Phi_2(z_2,\overline z_2)
\Phi_3(z_3,\overline z_3) \Phi_4(z_4,\overline z_4)} = 
\VEV{\prod_{i=1}^4 e^{i \alpha_i \phi(z_i,\overline z_i)}
(Q_+)^m (Q_-)^n}, 
\label{G4FFG}
\ee
where the charge neutrality condition is 
\be
\sum_{i=1}^4 \alpha_i + m\alpha_+ + n\alpha_- = 2\alpha_0. 
\label{zeromode}
\ee
In our model with $\alpha_0 =\frac12,\ \alpha_+ =2,\ \alpha_- =-1$ 
and $\alpha_i =\alpha_{1,2} =\frac12$ 
we choose $m=0$ and $n=1$. 
The correlation function (\ref{G4}) is, therefore, evaluated as 
\be
G^{(4)} 
= \VEV{\prod_{i=1}^4 e^{i \frac12 \phi(z_i,\overline z_i)} Q_-} 
= \prod_{i<j} |z_{ij}|^{\frac12} 
\int d^2u \prod_{i=1}^4 |z_i -u|^{-1}, 
\label{G4FF}
\ee
where $z_{ij}=z_i-z_j$. 
On the other hand, from the SL(2,{\bf C}) Ward identity \cite{BPZ}, 
the correlation function (\ref{G4}) can be written as 
\be
G^{(4)} =  F(x,\overline x) \prod_{i<j} |z_{ij}|^{\frac16}, 
\label{GF}
\ee
where $F$ is an arbitrary function of 
the SL(2,{\bf C}) invariant cross ratios 
\be
x ={z_{12}z_{34} \over z_{13}z_{24}}, \quad 
\overline x ={\overline z_{12} \overline z_{34} \over 
\overline z_{13} \overline z_{24}}. 
\ee
From Eqs.\ (\ref{G4FF}) and (\ref{GF}), 
by fixing $z_1=0,\ z_2=x,\ z_3=1,\ z_4=\infty$, 
$F(x,\overline x)$ can be evaluated and we obtain 
\be
G^{(4)} = |z_{13}z_{24}|^{\frac12} |x(1-x)|^{\frac12} 
{\textstyle I\left(-{1 \over2},-{1 \over2},-{1 \over2};x \right)}, 
\label{F}
\ee
where 
\be
I(a,b,c;x) = \int d^2 u \,|u|^{2a} |1-u|^{2b} |u-x|^{2c}. 
\label{myI}
\ee
The integral $I(a,b,c;x)$ can be transformed into 
a sum of squares of line integrals \cite{DF}
\be
I(a,b,c;x) = 
{\sin[\pi(a+b+c)] \sin(\pi b) \over \sin[\pi(a+c)]}|I_1(x)|^2
+ {\sin(\pi a) \sin(\pi c) \over \sin[\pi(a+c)]}|I_2(x)|^2, 
\label{refI}
\ee
where 
\ba
I_1(x) \A=\A \int_1^{\infty} du \, u^a (u-1)^b (u-x)^c 
\nonu
\A=\A
{\Gamma(-a-b-c-1) \,\Gamma(b+1) \over \Gamma (-a-c)}\,
F(-c,-a-b-c-1,-a-c;x), 
\label{I1} \\
I_2(x) \A=\A \int_0^x du \, u^a (1-u)^b (x-u)^c 
\nonu
\A=\A
{\Gamma(a+1) \,\Gamma(c+1) \over \Gamma (a+c+2)}\,
x^{a+c+1} \, F(-b,a+1,a+c+2;x). 
\label{I2}
\ea
The coefficients of $|I_1(x)|^2$ and $|I_2(x)|^2$ in 
Eq.\ (\ref{refI}) are determined by the monodoromy invariance 
of $I(a,b,c;x)$. 
We see that the result (\ref{F}) is naively an indeterminate form: 
\be
{\textstyle I \left(-{1 \over2},-{1 \over2},-{1 \over2};x \right)} 
= {1 \over \sin(-\pi)}\left[ - |I_1(x)|^2 + |I_2(x)|^2 \right] 
\sim {0 \over 0}, 
\label{0/0}
\ee
since $I_1(x) =I_2(x) =
\pi F \left({1 \over2},{1 \over2},1;x \right)$. 

To evaluate the above indeterminate form 
we now introduce a regularization procedure as follows: 
\ba
{\textstyle I \left(-{1 \over2},-{1 \over2},-{1 \over2};x \right)}
\A \equiv \A \lim_{a \rightarrow -{1 \over 2}}
{\textstyle I \left(a,-{1 \over2},-{1 \over2};x \right)}
\nonu
\A=\A
\left.{{d \over da}\left\{-\sin[\pi(a-1)] |I_1(x)|^2 
- \sin(\pi a) |I_2(x)|^2 \right\} \over 
{d \over da} \sin[\pi(a-{1 \over 2})]} 
\right|_{a =-{1 \over 2}}. 
\label{reg}
\ea
Note that $I_1(x)$ and $I_2(x)$ are 
the functions of the regularization parameter $a$, which are 
given by Eqs.\ (\ref{I1}) and (\ref{I2}) with $b=c=-\frac12$. 
Since $I_1(x) = I_2(x)$ when $a=-\frac12$, 
Eq.\ (\ref{reg}) becomes 
\be
{\textstyle I \left(-{1 \over2},-{1 \over2},-{1 \over2};x \right)} 
= {1 \over \pi} \left[ I_1(\overline x) 
\left\{ \left. {d \over da}\left(I_1(x) -I_2(x)\right) 
\right|_{a= -{1 \over 2}} \right\} 
+ (x \leftrightarrow \overline x) \right]. 
\label{reg2}
\ee
The factor of differential in the braces is evaluated as 
\ba
\left. {d \over da}\left(I_1(x) -I_2(x)\right)
\right|_{a=-{1 \over 2}} \A=\A -\pi \ln \left( \frac{x}{16} \right) 
{\textstyle F \left({1 \over 2},{1 \over 2},1;x \right)} - 2\pi M(x) 
\nonu
\A \equiv \A 2\pi \tilde F (x), 
\label{term1}
\ea
where
\be
M(x) = {1 \over \pi} \sum_{n=1}^\infty 
\left[{{\Gamma} \left(n+{1 \over 2}\right) \over n!}\right]^2 
\bigl[\psi(1) -\psi(n+1) -\psi({\textstyle {1 \over 2}}) 
+\psi(n + {\textstyle {1 \over 2}})\bigr] \, x^n 
\label{M}
\ee
and $\psi(x)$ is the digamma function. 
Anti-holomorphic part can be evaluated in the same way. 
Therefore we finally obtain 
\be
{\textstyle I \left(-\frac12, -\frac12, -\frac12; x \right)} 
= 2\pi \left[{\textstyle F \left(\frac12, \frac12, 1; x \right)} 
\tilde F (\overline x) 
+ (x \leftrightarrow \overline x) \right]. 
\label{Ifinal}
\ee
Notice that from Appendix C of Ref.\ \cite{HG} 
$\tilde F(x)$ satisfies the following relation 
\be
\tilde F (x) = \frac{\pi}{2} 
{\textstyle F \left(\frac12, \frac12, 1; 1-x \right)} 
= \int_0^{\pi \over 2}{d\theta \over \sqrt{1-(1-x)\sin^2 \theta}}. 
\ee
This is just the function $G(1-x)$ in Eq.\ (10) of 
Ref.\ \cite{LCFT1}, which is the origin of logarithm. 
Thus our result (\ref{F}) has logarithmic behaviour as 
\ba
G^{(4)} \A=\A \pi^2 
\left| z_{13} z_{24} \right|^{\frac12} 
\left| x(1-x) \right|^{\frac12} 
\nonu
\A\A \times 
\left[{\textstyle F \left(\frac12, \frac12, 1; x \right)} 
{\textstyle F \left(\frac12, \frac12, 1; 1- \overline x \right)} 
+ {\textstyle F \left(\frac12, \frac12, 1; \overline x \right)} 
{\textstyle F \left(\frac12, \frac12, 1; 1-x \right)} \right]. 
\ea
This agrees with that of Ref.\ \cite{LCFT1} up to overall constant, 
which was obtained by directly solving 
the hypergeometric differential equation. 

In the above procedure we performed an analytic continuation of 
the first parameter $a$ of the function $I(a,b,c;x)$. 
We can reproduce the same result by using the third parameter $c$ 
but not by the second one $b$. 
The fact depends on the choice of the two independent contours of 
integrals (\ref{I1}) and (\ref{I2}) since 
the coefficients of $|I_1(x)|^2$ and $|I_2(x)|^2$ in 
Eq.\ (\ref{refI}) are determined by monodoromy invariance 
of Eq.\ (\ref{refI}). 


\vspace{6mm}

\noindent
{\large{\bf Acknowledgements}}

The authors would like to thank Y. Tanii 
for careful reading of the manuscript. 
One of the authors (S.Y.) is also grateful to K. Hida and C.-B. Kim 
for useful discussions. 



\end{document}